# Broad Angle Negative Refraction in Lossless all Dielectric Multilayer Asymmetric Anisotropic Metamaterial

Ayed Al Sayem[1]*, M.R.C.Mahdy[2]*, and Md. Saifur Rahman[1]

[1]Dept. of EEE, Bangladesh University of Engineering and Technology, Dhaka, Bangladesh

[2] Department of Electrical and Computer Engineering, National University of Singapore, 4 Engineering Drive 3, Singapore

*Corresponding Authors' Emails: A0107276@nus.edu.sg

And ayedalsayem143@gmail.com

**Abstract:** In this article, it has been theoretically shown that broad angle negative refraction is possible with asymmetric anisotropic metamaterials constructed by only dielectrics or lossless semiconductors at the telecommunication and relative wavelength range. Though natural uniaxial materials can exhibit negative refraction, the maximum angle of negative refraction and critical incident angle lie in a very narrow range. This notable problem can be overcome by our proposed structure. In our structures, negative refraction originates from the highly asymmetric elliptical iso-frequency.This is artificially created by the rotated multilayer sub-wavelength dielectric/semiconductor stack, which act as an effective asymmetric anisotropic metamaterial.This negative refraction is achieved without using any negative permittivity materials such as metals. As we are using simple dielectrics, fabrication of such structures would be less complex than that of the metal based metamaterials. Our proposed ideas have been validated numerically and also by the full wave simulations considering both the effective medium approach and realistic structure model. This device might find some important applications in photonics and optoelectronics.

Negative refraction has been demonstrated from double negative metamaterial [1] to metal-dielectric multilayers [2-3], nanowire metamaterials [4-5], polar dielectrics [6], photonic crystals [7],antiferromagnets [8], hyper crystals [9, 10] and so on. Metal dielectric multilayers have been intensively used in constructing so called indefinite or hyperbolic metamaterials [2, 3, 11, 12] where the principal components of the dielectric tensor have opposite sign. Hyperbolic metamaterials possess the ability of negative refraction due to its hyperbolic iso-frequency [2-6, 13]. But as they are made from metals, loss in individual metal layer greatly reduces its performance and practical applications. Very recently, dielectric-dielectric/semiconductor sub-wavelength multilayers have been proposed as anisotropic elliptical metamaterial [13, 14] in their lossless frequency region. It has been already predicted that using such anisotropic metamaterial, wave-guiding performance would be better [14] in comparison with previous metamaterial based fibers [15]. Natural birefringence or uniaxial materials can also show negative refraction [16-20] but in a very narrow incident angle.Moreover these materials possess very narrow angle of refraction [16-20] which hinders their practical applications.

In this article, to the best of our knowledge, probably for the first time it is shown that broad angle negative refraction is possible by using an asymmetric anisotropic metamaterial made from loss less dielectric or semiconductor materials at the telecommunication and relative wavelength band. The angle of negative refraction and critical incident with the proposed lossless metamaterials can far exceed (better) those same properties observed with the natural uniaxial materials [16-20]. It can also be made tunable by changing the fill fraction and the rotation angle of the metamaterial, which defines its asymmetry. This negative refraction does not require any negative permittivity like others [4-5, 9, 10, 21] and there is no resonance like phenomena. So there is no possibility of loss as long as the dielectrics or semiconductors used are lossless in the respected frequency range of interest. Our ideas have been verified by full wave simulation considering exactly the realistic structure. Moreover,no ambiguity has arisedto verify our ideas based on effective medium theory. Effective medium theory may fail for certain conditions [22] those are not applicabe in ourproposed case. Also as we are using dielectrics or lossless semiconductors, the fabrication of our proposed structure should not be as complicated as metal based metamaterials [11].  We believe this lossless device with broad angle negative refraction may find some fundamental novel applications in photonics like in lensing, routing, beam splitting etc.

Fig. 1 shows the schematic of the proposed asymmetric anisotropic metamaterial. In the $(x',y',z')$ coordinate system the relative permittivity tensor of a general uniaxial anisotropic medium can be given by,

$$\boldsymbol{\varepsilon}' = \begin{bmatrix} \varepsilon_\| & 0 & 0 \\ 0 & \varepsilon_\| & 0 \\ 0 & 0 & \varepsilon_\perp \end{bmatrix} \quad (1)$$

Through rotation transformation around Y- axis at an angle φ with respect to X- axis, the permittivity tensor in the new coordinate system can be given by,

$$[\boldsymbol{\varepsilon}_{new}] = R_y [\varepsilon] R_y^T \quad (2)$$

Where, $R_y$ is the rotation matrix around the Y- axis and $R_y^T$ is the transposed matrix.

$$R_y = \begin{bmatrix} \cos\varphi & 0 & -\sin\varphi \\ 0 & 1 & 0 \\ \sin\varphi & 0 & \cos\varphi \end{bmatrix} \quad (3)$$

From (2) we get,

$$[\boldsymbol{\varepsilon}_{new}] = \begin{bmatrix} \varepsilon_{xx} & 0 & \varepsilon_{xz} \\ 0 & \varepsilon_{yy} & 0 \\ \varepsilon_{zx} & 0 & \varepsilon_{zz} \end{bmatrix} = \begin{bmatrix} \varepsilon_\| \cos^2\varphi + \varepsilon_\perp \sin^2\varphi & 0 & (\varepsilon_\| - \varepsilon_\perp)\sin\varphi\cos\varphi \\ 0 & \varepsilon_\| & 0 \\ (\varepsilon_\| - \varepsilon_\perp)\sin\varphi\cos\varphi & 0 & \varepsilon_\|\sin^2\varphi + \varepsilon_\perp \cos^2\varphi \end{bmatrix} \quad (4)$$

In Fig. 1 two Cartesian systems have been used, $(x',y',z')$ before rotation and $(x,y,z)$ after rotation. ATM polarized electromagnetic wave (Magnetic field polarized along y axis) is incident from the ambient medium (air) to the structure, propagating in x-z plane forming an angle $\theta$ with z axis. We have taken the space-time dependence of fields as $e^{-i(\omega t - k_z z - k_x x)}$.

Now considering similar permeability tensor, in the new (x, y, z) coordinate system after rotation, the normal component of the wave vector is derived by rigorous calculation as (cf. supplementary information for detail analysis considering both permittivity and permeability tensor),

$$k_z = \frac{\varepsilon_{xz}\tilde{\mu}_{yy} k_x \pm \sqrt{\tilde{\mu}_{yy}^2 (\varepsilon_{xz}^2 - \varepsilon_{xx}\varepsilon_{zz})k_x^2 - k_0^2 \varepsilon_{zz}\tilde{\mu}_{yy}(\varepsilon_{zx}\varepsilon_{xz} - \varepsilon_{zz}\varepsilon_{xx})}}{\varepsilon_{zz}\tilde{\mu}_{yy}} \quad (5)$$

Where $k_x$ is the tangential component of wave vector or momentum, which is defined by $k_o \sin\theta$, where $k_o = \frac{2\pi}{\lambda}$ is the wave-number in free space and $\lambda$ is the wavelength of incident EM wave. For isotropic magnetic case, $\tilde{\mu}_{yy} = \frac{1}{\mu}$ ($\mu$ is the relative permeability) and for symmetric case $\varepsilon_{zx} = \varepsilon_{xz}$, so equation (5) reduces to,

$$k_z = \frac{\varepsilon_{xz} k_x \pm \sqrt{(\varepsilon_{xz}^2 - \varepsilon_{xx}\varepsilon_{zz})k_x^2 - k_o^2 \varepsilon_{zz} \mu(\varepsilon_{zx}\varepsilon_{xz} - \varepsilon_{zz}\varepsilon_{xx})}}{\varepsilon_{zz}} \tag{6}$$

In equation (6) the (+) and the (-) signs are for the normal wave vector components propagating in the downward and upward direction respectively. For TM (p) polarization, **H** field is confined along the y direction. The time-average poynting vector can be calculated as $\langle \mathbf{S}_{Amm} \rangle = \frac{1}{2} Re[\mathbf{E} \times \mathbf{H}^*]$. The x and z component of the time average poynting vector have been derived and they are given by,

$$\langle S_{Amm\,x} \rangle = \frac{1}{2} \frac{|H_y|^2}{\omega \varepsilon_0} Re\left(\frac{\varepsilon_{xx} k_x - \varepsilon_{zx} k_z}{\varepsilon_{zz}\varepsilon_{xx} - \varepsilon_{zx}\varepsilon_{xz}}\right) \tag{7}$$

$$\langle S_{Amm\,z} \rangle = \frac{1}{2} \frac{|H_y|^2}{\omega \varepsilon_0} Re\left(\frac{k_z \varepsilon_{zz} - k_x \varepsilon_{xz}}{\varepsilon_{zz}\varepsilon_{xx} - \varepsilon_{zx}\varepsilon_{xz}}\right) \tag{8}$$

The angle of refraction for the wave vector can be obtained as,

$$\tan\theta_{r,k} = \frac{Re(k_x)}{Re(k_z)} \tag{9}$$

And the angle of refraction for the poynting vector can be obtained as,

$$\tan\theta_{r,s} = \frac{\langle S_{Amm\,x} \rangle}{\langle S_{Amm\,z} \rangle} = \frac{Re(\varepsilon_{xx} k_x - \varepsilon_{zx} k_z)}{Re(k_z \varepsilon_{zz} - k_x \varepsilon_{xz})} \tag{10}$$

(cf. supplementary information for details of theoretical derivations). For $\varphi = 0$ ($x = x', y = y', z = z'$), $\varepsilon_{zx} = \varepsilon_{xz} = 0$ and equation (10) reduces to the most familiar form [4, 5] for uniaxial metamaterials,

$$\tan\theta_{r,k} = \frac{Re(\varepsilon_{xx} k_x)}{Re(k_z \varepsilon_{zz})} \tag{11}$$

For hyperbolic metamaterials: $\varepsilon_{xx}\varepsilon_{zz} < 0$. As a result, negative refraction can be achieved [4, 5]. To construct a hyperbolic metamaterial and to achieve the condition, $\varepsilon_{xx}\varepsilon_{zz} < 0$, one must has to use metal in the structure (metal-dielectric multilayer planer structure [2, 3, 11] or metallic nano-rod in dielectric host [4, 5]). Metals are usually lossy and hinder the practical applications of negative refraction.

Now when constructing an anisotropic metamaterial only with sub-wavelength dielectrics, all the tensor components in equation (1) is always positive. Therefore for non-rotated case, $\varphi =$

$0$ $(x = x', y = y', z = z')$, $\varepsilon_{xx} = \varepsilon_x$, $\varepsilon_{zz} = \varepsilon_z$, $\varepsilon_{xz} = \varepsilon_{zx} = 0$ and right side of equation (10) is always positive (same is true for φ=±90°) and negative refraction can not be achieved. But interesting phenomena can happen when the structure is rotated. From equation (10), depending on the values of $\varepsilon_{xx}, \varepsilon_{xz} = \varepsilon_{zx}$ and $\varepsilon_{zz}$, right side of equation (10) can be negative. As a result, negative refraction can be achieved. This can be easily visualized by the equi-freqeucney contour in wave vector space of the proposed metamaterial.

Without using arbitrary values of $\varepsilon_{xx}, \varepsilon_{xz}$ and $\varepsilon_{zz}$, we are more interested in practical realization of deep sub-wavelength all dielectric multilayer structure. This multilayer structure is made by alternatively repeating two different dielectric materials with permittivity $\varepsilon_{d1}$ and $\varepsilon_{d2}$ and thickness $t_1$ and $t_2$. From effective medium theory, the values of permittivity tensor in equation (1) can be given by,

$$\varepsilon_\parallel = \varepsilon_x = \varepsilon_y = f\varepsilon_{d1} + (1-f)\varepsilon_{d2} \tag{12a}$$

$$\varepsilon_\perp = \varepsilon_z = \left(\frac{f}{\varepsilon_{d1}} + \frac{1-f}{\varepsilon_{d2}}\right)^{-1} \tag{12b}$$

where $f$ is fill fraction given by, $f = \frac{t_1}{t_1+t_2}$.

Fig. 2a shows the effective permittivity ($\varepsilon_\parallel$ and $\varepsilon_\perp$) of Air-Silicon and SiO$_2$-silicon multilayer metamaterial as a function of fill factor. Unlike the case metal dielectric multilayers [2, 3, 11], both $\varepsilon_\parallel$ and $\varepsilon_\perp$ are confined to values between $\varepsilon_{d1}$ and $\varepsilon_{d2}$. Fig. 2b shows the effective permittivity ($\varepsilon_{xx}, \varepsilon_{xz}, \varepsilon_{yy}$ and $\varepsilon_{zz}$) as a function of rotation angle φ around Y axis for the Air-Silicon and SiO$_2$-Silicon multilayer metamaterial respectively.

Fig. 3 shows the equi-frequency contour in k-space of the proposed asymmetric anisotropic metamaterial. Here as an example, we haveused air-silicon multilayer metamaterial. Fig.3a (φ=0°), fig. 3d, (φ=90°) represents the equi-frequency contour for the non-rotated case and fig. fig. 3b (φ=45°) and fig. 3c (φ=-45°) represents the equi-frequency contour for the rotated case. Refraction of a wave incident from one medium to another can be easily visualized by the equi-frequency contour in the wave vector space. As we know, the group velocity can be expressed as, $\boldsymbol{v_g} = \frac{d\omega}{d\boldsymbol{k}}$ which coincides with the direction of the time averaged Poynting vector or the direction of energy flow or power flow. So the direction of group velocity or energy flow at any point k of the equi-frequency canbe determined by drawing a normal line to the tangent line on that point (indicated by the green and red arrows in fig 3a-d). Of course the tangential momentum has to be conserved at the interface. The blue circle represents the equi-frequency contour of air and the blue lines are drawn to specify different tangential momentum. Form fig. 3a (rotation angle, $\varphi=0^o$) it can be visualized that negative refraction would not be possible for the non-rotated case and also for rotation angle, $\varphi=\pm 90^o$ (fig. 3d), which coincides with the mathematical explanation above. But for any other rotation angle, negative refraction will be possible as indicated by fig. 3b and 3c.

The directions of wave and energy propagation for the extraordinary mode (p polarization) in birefringentor uniaxial medium do not coincide [20]. For non-rotated case ($\varphi=0^o$, $\pm 90^o$), the group velocity or power flow direction always remains on different side of the surface normal indicating positive refraction. But for rotated case and due to highly elliptical equi-frequency, the refracted wave can occur on the same side of the surface normal [18] indicating negative refraction (fig. 3b, 3c). In case of hyperbolic metamaterials, negative refraction occurs due to the hyperbolic iso-frequency. To achieve this hyperbolic iso-frequency,$\varepsilon_{xx}\varepsilon_{zz} < 0$ this condition (equation 11) must be fulfilled [4, 5]. It indicates that one must use metallic materials for constructing hyperbolic metamaterials. In case of elliptical metamaterials, the case which is considered here, negative refraction occurs due to the highly elliptical iso-frequency and only for the case when optical axis is rotated at certain angle to the interface normal.Though all angle negative refraction will not be possible, broad angle negative refraction can be easily achieved. As lossless materials are used to construct the metamaterial, there will be no absorption and high transmission will be feasible. In addition, as the materials are almost non-dispersive in the considered frequency range, broadband negative refraction will be possible especially in the telecommunication wavelength bands.

Fig. 4a shows that angle of refraction as a function of incident angle for different combination of materials (used for constructing the asymmetric anisotropic metamaterials) with rotation angle, $\varphi=30^o$. Both maximum negative refraction angle and critical angle increases with the difference between the permittivity values of individual layer of the metamaterial. Also maximum negative refraction is achieved for incident angle, $\theta=0^o$. Fig. 4b shows the 2d map of the angle of refraction as a function of incident angle and angle of rotation. As the materials in the considered frequency (telecommunication wavelength, λ=1.55µm) range is loss less, probability of the absorption loss is extremely low. Fig. 5a and 5b show the 2d map of transmission of a $SiO_2$-Silicon and air-Silicon asymmetric anisotropic metamaterial respectively calculated by general 4×4 transfer matrix [23-25]. It can be visualized that very high transmission can be achieved for a broad range of incident and rotation angle.

Table I shows the maximum refraction and critical incident angle for different combinations of materials used for constructing the asymmetric anisotropic metamaterial along with the natural uniaxial anisotropic materials such as $YVO_4$, $CaCO_3$, a-BBO [26]. It is clearly seen that both maximum angle of negative refraction and maximum critical incident angle with the proposed lossless metamaterials can far exceed (better) those same properties observed with the natural uniaxial crystals.

**Table I: Maximum Angle of Negative Refraction and critical angle for different combination of materials along with natural ones**

| Combination (Metamaterial or Natural) | Maximum Angle of Negative Refraction (Degree) | Maximum Critical Incident Angle (Degree) |
|---|---|---|
| Air-$SiO_2$ | 3.78 | 4.6 |
| Air-GaAs | 32.82 | 90 |
| Air-Silicon | 34.06 | 90 |
| $SiO_2$-GaAs | 18.23 | 45.9 |
| $SiO_2$-Silicon | 19.39 | 51 |
| $SiO_2$-Germanium* | 27.54 | 90 |
| $CaCO_3$ (natural) [26] | 5.7714 | 10 |
| $YVO_4$ (natural) [26] | 5.8842 | 13 |
| a-BBO (natural) [26] | 4.4859 | 8 |

Wavelength, λ=1.55µm
*not fully lossless

To verify our theoretical proposal, full wave simulations have been performed. Fig. 6 demonstrates the map of the transverse magnetic field component $H_y$ when a TM (p) polarized Gaussian beam ($\lambda=1.55\mu m$) is incident from air to Air-Silicon Multilayer metamaterial. The structure is rotated at an angle $\varphi \approx 59^0$. In fig. 6a and 6b, wave is incident at an angle $\theta=0^0$ and in fig. 6c and 6d wave is incident at an angle $\theta=25^0$. In fig. 6a and 6c, real multilayer structure has been considered where each layer thickness is 100nm. But in fig. 6b and 6d, effective medium description has been used to characterize the metamaterial. The simulation results considering the real structure and the effective medium approximation are in quite good agreement. At both incident angles, negative refraction of light is evident, which coincides with our theoretical predictions. More simulation figures (both real structure and effective medium description) considering different dielectric materials and different rotation angle are available in the supplementary.

In summary, the possibility of broad angle negative refraction by asymmetric anisotropic metamaterial using only nonmetallic and lossless materials has been theoretically demonstrated in the telecommunication wavelength band. This lossless metamaterial can have superior performance over natural uniaxial materials and also over metal based highly absorbing metamaterials. As dielectrics are less dispersive in the considered frequency range, the proposed metamaterial is expected to be broadband. Moreover, because of the absence of metal in the proposed multilayer metamaterial, its fabrication is expected to be much simpler than that of the metal based multilayer metamaterials. Our ideas have been validated by numerical results and by full wave simulations considering both effective medium and realistic structure. We believe this work may find important applications in photonics [27, 28] like in lensing, beam splitting etc.and also in new semiconductor research [29].

**Supplementary article:** Supplementary online article contains detail derivations and the details of the numerical calculations. Also some additional full wave simulation results are available in the supplement.

**Figures and Captions**

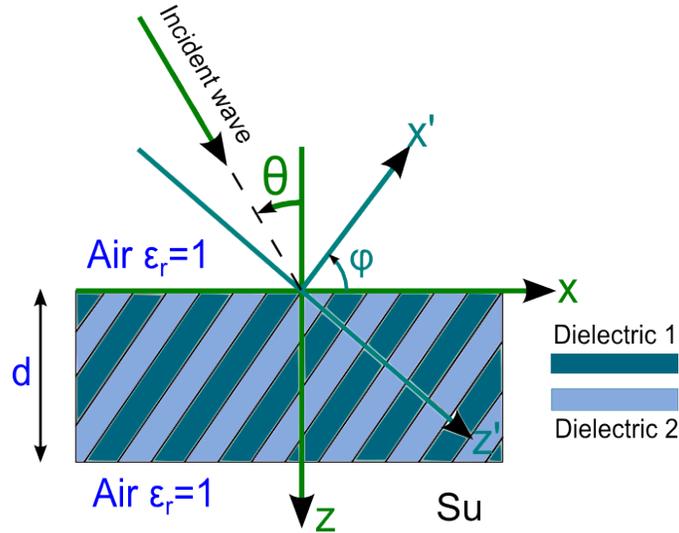

Fig. 1: Schematic of the proposed asymmetric anisotropic multilayer metamaterial. Individual slab interface lies along the $x'$ axis. The structure is rotated around y axis at an angle φ with respect to x axis. d is the thickness of the structure. A TM (p) polarized electromagnetic wave (magnetic field polarized along y axis) is incident at angle θ with respect to the z axis.

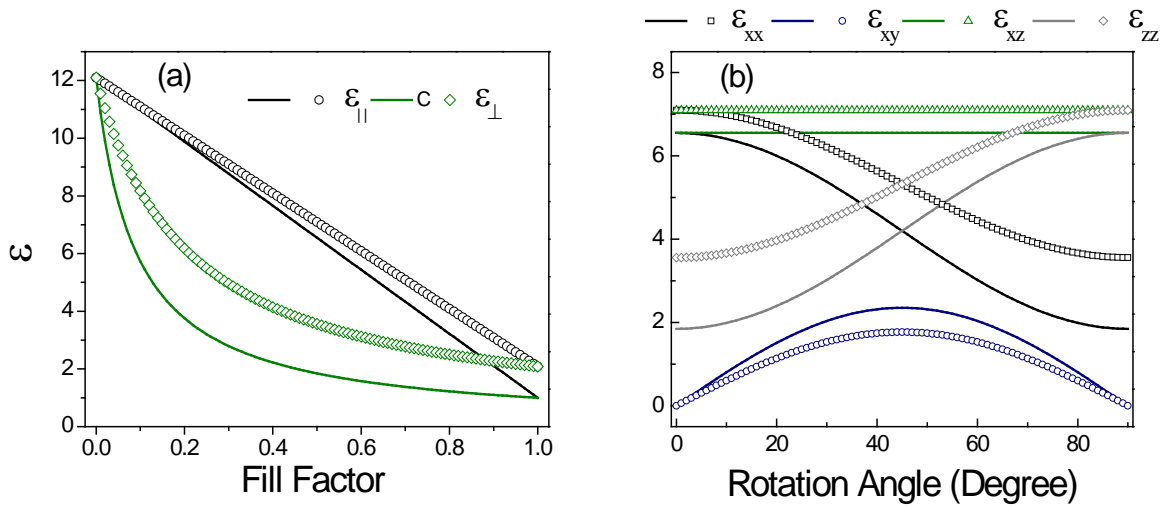

Fig. 2: (a) Effective permittivity (tangential $\varepsilon_{\parallel}$ and perpendicular $\varepsilon_{\perp}$) of air-silicon and SiO$_2$-silicon multilayer metamaterial as a function of fill factor. (b) Effective permittivity $\varepsilon_{xx}$, $\varepsilon_{xz}$, $\varepsilon_{yy}$ and $\varepsilon_{zz}$ of the asymmetric anisotropic metamaterial as a function of rotation angle around the y axis. Solid lines and symbols are for Air-Silicon and SiO$_2$-Silicon based metamaterial respectively.

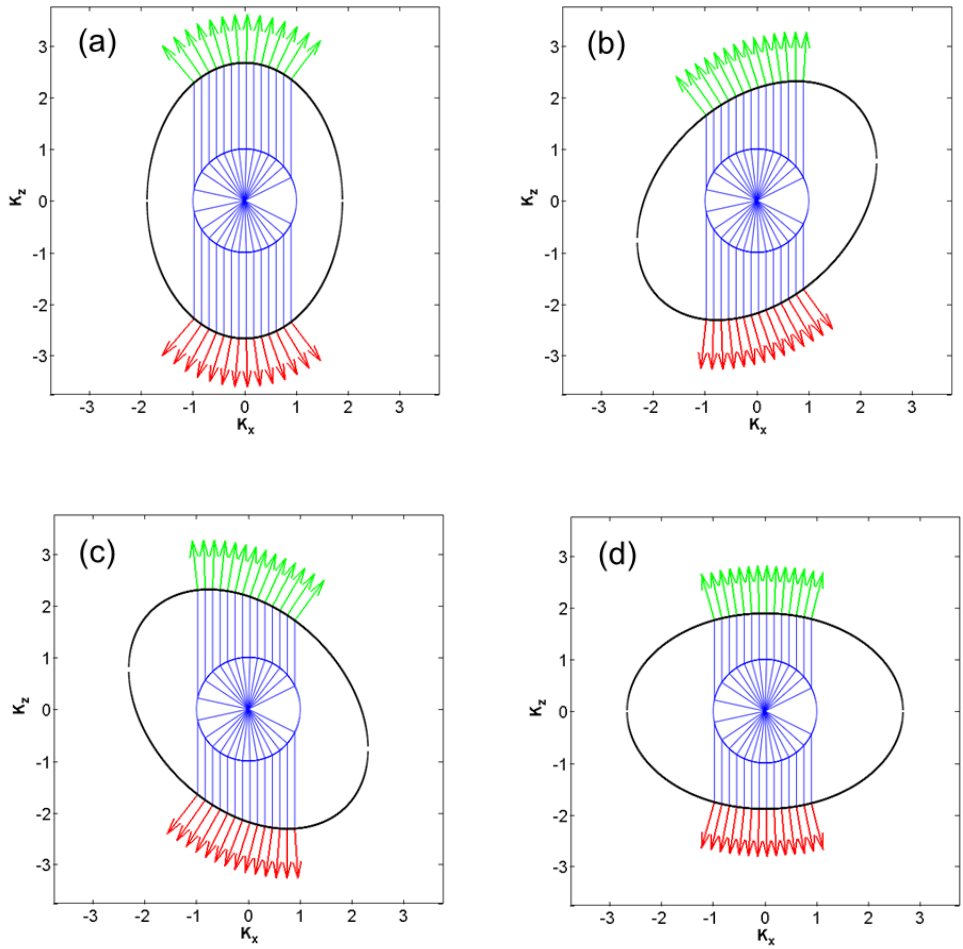

Fig. 3: The equi-frequency contour in k-space of the proposed asymmetric anisotropic metamaterial with rotation angle around y axis (a) $\varphi=0^o$ (b) $\varphi=45^o$ (c) $\varphi=-45^o$ (d) $\varphi=90^o$ (In this case, we are using Air-Silicon metamaterials). The blue circle represents the equi-frequency contour of air. The green and red arrows indicate the power flow or group velocity direction. Blue lines are drawn to indicate specific tangential ($k_x$) wave vector for specific group velocity direction.

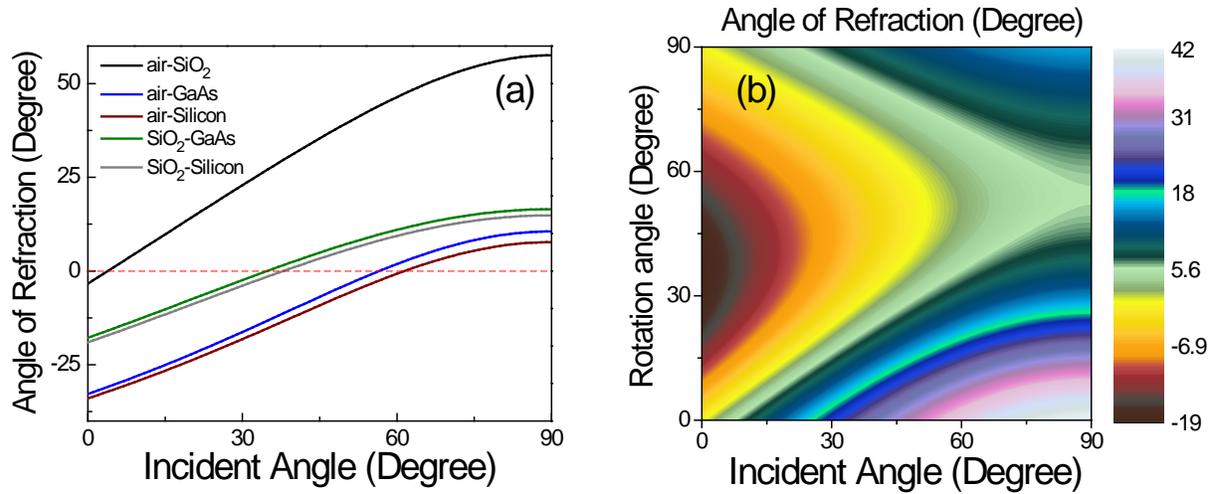

Fig. 4: (a) Angle of refraction of the asymmetric anisotropic metamaterial as a function of incident angle for different combination of materials (Air-SiO$_2$, Air-GaAs, Air-Silicon, SiO$_2$-GaAs, SiO$_2$-Silicon). (b) 2D map of refraction angle as a function of the incident angle and the rotation angle for Air-Silicon based asymmetric anisotropic metamaterial.

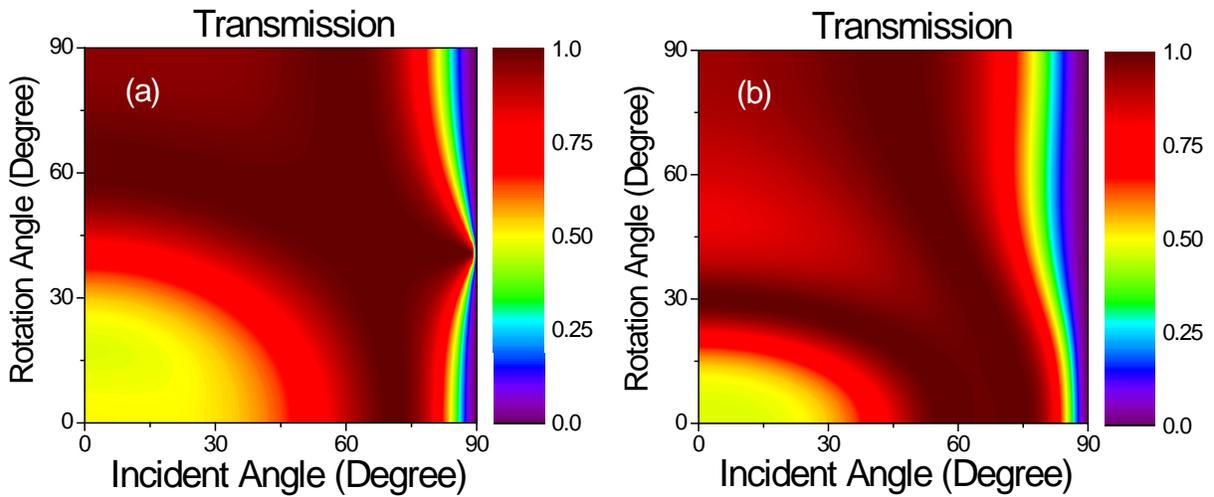

Fig. 5: 2D map of Transmission co-efficient of p-polarized wave (wavelength, $\lambda=1.55\mu m$) as a function of incident angle and rotation angle calculated by 4×4 transfer matrix method for (a) SiO$_2$-Silicon (b) Air-Silicon asymmetric anisotropic metamaterial.

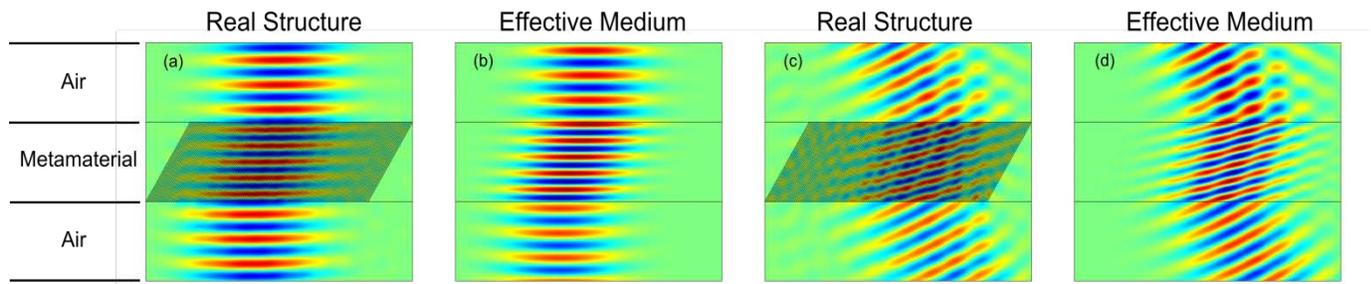

Fig. 6: Full wave numerical simulations (Map of the transverse magnetic field component, $H_y$) demonstrate negative refraction of a monochromatic ($\lambda=1.55\mu m$) TM-polarized Gaussian beam. The beam is incident from air to Air-Silicon multilayer metamaterial with incident angle (a, b) $\theta=0^0$ (c, d) $\theta=25^0$. The structure is rotated at an angle $\varphi \approx 59^0$ and each layer thickness is 100nm. (a) and (c) represent the field components for the real rotated multilayer metamaterial structure. (b) and (d) represent the field components with its effective medium description (instead of the real metamaterial structure).